\documentclass[twocolumn,superscriptaddress,floatfix,aps]{revtex4}%
\usepackage{eurosym}
\usepackage{graphicx}
\usepackage{dcolumn}
\usepackage{ulem}
\usepackage{textcomp}
\usepackage{amssymb}
\usepackage{amsmath}
\usepackage{color}
\usepackage{footnote}
\usepackage{microtype}
\usepackage{amsfonts}%
\begin{document}
\title{Renormalized q-dependent Spin Susceptibility by inverting the
Random Phase Approximation: Implications for quantitative assessment of the
role of spin fluctuations in 2D Ising superconductor NbSe$_{2}$}
\author{Suvadip Das}
\author{Igor I. Mazin}
\email[email: ]{imazin2@gmu.edu}
\affiliation{Department of Physics and Astronomy and Quantum Science and Engineering
Center, George Mason University - Fairfax, VA, USA }

\date{\today}

\begin{abstract}
Accurate determination of the full
momentum-dependent spin susceptibility $\chi(\mathbf{q}) $ is very important for the description of magnetism and superconductivity. While {\it in principle} the formalism for calculating  $\chi(\mathbf{q})$ in the linear response density functional theory (DFT) is well established, hardly any publicly available code includes this capability. Here, we describe an alternative way to calculate  the static  $\chi(\mathbf{q})$,
which can be applied to most common DFT codes without additional programming. The method combined standard fixed-spin-moment calculations of $\chi(\mathbf{0}) $ with direct calculations of the energy of spin spirals stabilized by an artificial Hubbard interaction. From these calculations, $\chi_{DFT}(\mathbf{q} )$ can be extracted by inverting the RPA formula. We apply this recipe to the recently discovered Ising superconductivity in NbSe$_2$ monolayer, one of the most exciting findings in 
superconductivity in recent years. It was proposed that spin fluctuations may strongly affect the parity of the order parameter. Previous estimates suggested proximity to ferromagnetism, $i.e.$,  $\chi(\mathbf{q})$ peaked at $\mathbf{q}=0$. We find that the
structure of spin fluctuations is more complicated, with the fluctuation spectrum sharply peaked at $\mathbf{q}\approx (0.2,0)$. Such a spectrum would change the interband pairing interaction  and considerably affect the superconducting state.
\end{abstract}
\maketitle

\section{Introduction}

Knowledge of the full momentum-dependent spin susceptibility $\chi
(\mathbf{q)}$ is very important in condensed-matter physics~~\cite{Moriya}. In
particular, it is a key parameter in the theory of spin-fluctuation induced
superconductivity~\cite{Berk,Fay}, which has been the subject of intensive 
research in the last few decades. Moreover, it was recently emphasized that
spin fluctuations may play a crucial role in determining the superconducting
state property and pairing symmetry even when spin-fluctuations provide a
subdominant pairing interaction. This was argued to be the
case~\cite{Darshana,Mak_arxiv} in one of the most exotic recent discoveries in
superconductivity, the so-called Ising superconductivity in NbSe$_{2}$
monolayers~\cite{Mak,Khodas}.

Density functional theory (DFT) provides a good starting point, even though in
itinerant magnets it overestimates the tendency to magnetism~\cite{Moriya}.
Unfortunately, calculation of the full spin susceptibility, while conceptually
straightforward in DFT, is involved, and, most importantly, such capabilities
are not included in the common DFT software packages.~\cite{VASP,elk,FLEUR,wien2k} Relatively few
publications report such calculations,
~\cite{Savrasov,Hardy,Massidda,Schilfgaarde,Kutepov,Antropov}, and they are
all based on custom-built programs. On the other hand, essentially all popular
DFT packages include the capability for fixed spin moment calculations, which
provide the exact value for the \textit{uniform} DFT susceptibility
$\chi_{DFT}(0)$ at a low computational cost. Comparing thus calculated
$\chi_{DFT}(0)$ with the \textit{unrenormalized} one-electron susceptibility
$\chi_{DFT}^{(0)}(0)\equiv N_{F}(0),$ where $N_{F}(0)$ is the
density of states per spin at the Fermi level (here and throughout the paper
we are using the atomic units convention where the Bohr magneton is chosen to be 1), one can determine the
so-called Stoner factor that describes the effect of electron-electron
interaction within DFT on the spin susceptibility:%
\begin{equation}
\chi_{DFT}(0)=\frac{\chi_{DFT}^{(0)}(0)}{1-I\chi_{DFT}^{(0)}(0)}
\end{equation}
Note that, apart from the Umklapp processes this expression is exact in DFT
(albeit not in the many-body theory~\cite{Ron}).

While cases are known when the Stoner factor has a non-negligible $\mathbf{q}$
dependence~\cite{David}, these are uncommon and usually setting $I$ to a
\textbf{q}-independent constant is a good approximation. Moreover, following
Moriya's Self-Consistent Renormalization Theory~\cite{Moriya} one can account
for the effect of spin-fluctuations reducing the tendency to magnetism by
replacing $I$ with an effective, reduced interaction $I_{eff}=\alpha I,$
$\alpha<1.$ This approach is sometimes called \textit{Reduced Stoner Theory}
(RST)~\cite{Ortenzi}.

On the other hand, one can go beyond DFT by adding a local Coulomb
interaction, $U_{eff}$, through the so-called LDA+U method~\cite{Dudarev}. In fact, it
\textit{increases }the tendency to magnetism, rather than decreasing it, as would be required for a better agreement with the experiment in
itinerant magnets, but, as we
will discuss below, gives us a formal tool to calculate $\chi_{DFT}%
(\mathbf{q)}$ without engaging the linear response theory. It was shown that, to
a good approximation, this method adds an addition contribution to $I,$ namely
$\kappa U_{eff},$ where the coefficient $\kappa$ is material dependent and
reflects the orbital composition of the states near the Fermi
level~\cite{Petukhov}.

In this paper, we propose a simplified way to estimate $\chi_{DFT}%
(\mathbf{q}),$ and, by using RST, the fluctuation-corrected $\chi
(\mathbf{q}),$ without doing full linear response calculations. The only
prerequisite is a DFT package that allows the LDA+U
extension (essentially all modern tools do) and spin-spiral calculations (most
popular packages such as VASP~\cite{VASP}, ELK~\cite{elk}, FLEUR~\cite{FLEUR}, WIEN2k~\cite{wien2k}
have this capability as well). We further illustrate this approach by
calculating $\chi_{DFT}(\mathbf{q})$ for the Ising superconductor NbSe$_{2}$ monolayer.

The paper is organized as follows. First, we present the general theory of the
spin susceptibility in the Random Phase Approximation (RPA), which is exact in
both DFT and LDA+U. Second, we describe the algorithmic steps to extract
$\chi_{DFT}(\mathbf{q})$ for a given \textbf{q}. Finally, we present
comprehensive results and relevant discussions for our system of interest,
NbSe$_{2}$ monolayers.

\section{General Theory}

\subsection{Spin susceptibility in DFT and beyond}

The most general definition of spin susceptibility is given in the real space
\begin{equation}
\chi^{-1}(\mathbf{r,r}^{\prime})=\frac{\delta^{2}E}{\delta m(\mathbf{r)}\delta
m\mathbf{(r}^{\prime})}, \label{1}%
\end{equation}
where $E$ is the total energy of the system. In DFT, it can be written exactly
as
\begin{equation}
E=E_{1}+E_{xc}+E_{ns}%
\end{equation}
where $E_{1}$ is the one-electron energy (sum of the DFT eigenenergies for
all occupied states), $E_{xc}$ is the exchange-correlation energy, usually
computed in either the Local Density Approximation (LDA) or in the Generalized
Gradient Approximation (GGA)~\cite{Ceperley,GGA}, and $E_{ns}$ does not depend
on the spin density. One can then introduce%
\begin{align}
\chi_{0}^{-1}(\mathbf{r,r}^{\prime})  &  =\frac{\delta^{2}E_{1}}{\delta
m(\mathbf{r)}\delta m\mathbf{(r}^{\prime})}\\
I(\mathbf{r,r}^{\prime})  &  =-\frac{\delta^{2}E_{xc}}{\delta m(\mathbf{r)}%
\delta m\mathbf{(r}^{\prime})}\\
\chi^{-1}_{DFT}(\mathbf{r,r}^{\prime})  &  =\chi_{0}^{-1}(\mathbf{r,r}^{\prime
})-I(\mathbf{r,r}^{\prime})
\end{align}

Upon Fourier transform, neglecting the Umklapp local field
effects~~\cite{Igor,Ron},%
\begin{equation}
\chi^{-1}_{DFT}(\mathbf{q})=\chi_{0}^{-1}(\mathbf{q})-I
\end{equation}
where, as discussed in the Introduction, the \textbf{q} dependence of $I$ is
neglected. Consequently, the RPA approximation ~\cite{Pines},%
\begin{equation}
\chi_{DFT}(\mathbf{q})=\frac{\chi_{0}(\mathbf{q})}{1-I\chi_{0}(\mathbf{q})}%
\end{equation}
is exact. The \textquotedblleft fixed spin moment\textquotedblright\ (FSM)
method, applicable for $\mathbf{q}=0,$ utilizes Eq. \ref{1} directly:%
\begin{equation}
\chi_{DFT}^{-1}(0)=\frac{\delta^{2}E}{\delta M^{2}}=\chi_{0}^{-1}(0)-I
\label{dft}%
\end{equation}
where $M$ is the total magnetization. Modifications described above come as
additional terms in this formula%
\begin{equation}
\chi_{RST}^{-1}(0)=\frac{\delta^{2}E}{\delta M^{2}}=\chi_{0}^{-1}(0)-\alpha I
\label{eq:eq1}%
\end{equation}
where $\alpha$ can be determined from comparison with the experiment, and%
\begin{equation}
\chi_{LDA+U}^{-1}(\mathbf{q})=\frac{\delta^{2}E}{\delta M^{2}}=\chi_{DFT}%
^{-1}(\mathbf{q})-\kappa U_{eff}, \label{U}%
\end{equation}
where $U_{eff}=\mathrm{(U-J)},$ as defined in Refs. ~\cite{Dudarev,Petukhov}.

In principle, one can apply the FSM recipe to finite wave vectors, but very
few codes allow frozen spin-wave calculations with fixed amplitude,
%~\cite{FLEUR, elk}% DO THEY?
 and in those that do, it is cumbersome and time-consuming.
Alternatively, one can use LDA+U and Eq. \ref{U} to extract $\chi_{DFT}%
^{-1}(\mathbf{q})$ from the instability condition:%
\begin{equation}
\chi_{DFT}^{-1}(\mathbf{q})-\kappa U_{eff}=0,\label{Uc}%
\end{equation}
The recipe is then to vary $U_{eff}$ until the nonmagnetic solution becomes
unstable. As mentioned, $\kappa$ can be determined by applying Eq. \ref{U} at
$\mathbf{q}=0$ and comparing with standard FSM calculations.

One \textit{caveat} is in place. While the above equations deal with
infinitesimally small magnetic moments, in reality meta-magnetic states with
two metastable solutions, at $M=0$ and at a finite $M$ may exist. The way to deal with this situation is to always start calculations from a very small moment, making sure that even if the $M=0$ is not the ground state, the program does not leave this minimum as long as it remains metastable.

\subsection{Enhancement of Stoner exchange using DFT+U in the spin
susceptibility}

The instability of the paramagnetic ground state is dictated by the Stoner
criterion for ferromagnetism ~\cite{Blundell}, indicative of strong
electron-electron interactions in the system. The latter can be tuned, in a simple way, by including additional on-site interactions in form
of the standard Hubbard model in the static mean field approximation, known as ``LDA+U'' (or, more correctly, DFT+U) method.
While DFT underestimates the tendency to magnetism in strongly localized
electronic systems, DFT+U compensates for that by incorporating the orbital-selective Hubbard interaction of the strongly
localized electrons. In our study, we use the spherically averaged and
rotationally invariant LDA+U methodology proposed by Dudarev {\it 
et al}.~\cite{Dudarev}.
\begin{align}
E_{LSDA+U}  &  =E_{LSDA}+\frac{({U}-{J})}{2}\sum_{\sigma}(n_{m,\sigma
}-n_{m,\sigma}^{2})\\
&  =\frac{({U}-{J})}{2}\sum_{\sigma}\mathrm{Tr}(\rho_{\sigma})-\mathrm{Tr}(\rho_{\sigma}%
\rho_{\sigma})
\end{align}
where ${U}$ and ${J}$ are the spherically averaged Hubbard repulsion and
intra-atomic exchange for electrons with the given angular momentum
$\mathit{l}$, $n_{m,\sigma}$ is the occupation number of the $m$th orbital, and
$\sigma$ is the spin index. The magnetic interactions can then be
efficiently tuned by adding an effective Hubbard parameter
${U_{eff}={(U-J)}}$ as shown by Petukhov
$et$ $al$~\cite{Petukhov}. Note that the orbital selective contribution of
the effective Hubbard term ${U_{eff}={(U-J)}}$ plays an important role in
determining the Stoner factor within the Density Functional Theory framework.
Utilizing DFT, the Stoner parameter $I$ can be expressed as ${I=-2\partial
^{2}E_{xc}/\partial M^{2}}$, the second derivative of the exchange-correlation
energy with respect to the total magnetic moment. The paramagnetic ground
state becomes unstable when ${N_{F}I\geq1}$. Upon incorporation of the orbital
dependent Hubbard $U$ parameter, there is an enhancement of the Stoner factor
compared to DFT. Within the \textquotedblleft fully localized
limit\textquotedblright\ (FLL), the correction to the total energy due to the
$DFT+U$ can be written as~\cite{Petukhov}
\begin{equation}
{\Delta E_{LDA+U}^{FLL}=-\frac{(U-J)}{2}\sum_{\sigma}\mathrm{Tr}({\rho}^{\sigma}.{\rho
}^{\sigma})-(2l+1)n^{\sigma}}%
\end{equation}
This results in an additional contribution to the Stoner parameter
\begin{equation}
{\Delta I=\frac{(U-J)}{N_{F}^{2}}}\mathrm{Tr}{(D\cdot D)} \label{eq:eq3}%
\end{equation}
where ${D_{mm^{^{\prime}}}=-\pi^{-1}ImG_{mm^{^{\prime}}}(E_{F})}$ is
proportional to the imaginary part of the corresponding Green's function. This
additional contribution is proportional to the effective Hubbard term
${U_{eff}={(U-J)}}$, and to the factor, $Tr(D\cdot D)$, which depends on the
orbital composition of the bands at the Fermi level, usually can be safely
chosen to be a $\mathbf{q}-$independent constant, for a given system, thus the
additional term can be simply written as ${\Delta I=\kappa U}_{eff}$.

\subsection{Spin-spiral calculations}

It was pointed out about 30 years ago by L. M. Sandratskii~\cite{Sandratskii,
Sandratskii1, Sandratskii2} that when solving a single-particle Scr\"{o}dinger
equation in a spiral magnetic field (not necessarily commensurate with the
periodicity of the charge potential) a generalized Bloch theorem can be
derived, along the following lines:  

Let us assume that the spin density in a given unit cell is related to that in
all other unit cells as below:
\begin{equation}
\mathbf{M}(\mathbf{r}+\mathbf{R})=%
\begin{pmatrix}
M_{x}(\mathbf{r})\cos(\mathbf{q\cdot R})-M_{y}(\mathbf{r})\sin(\mathbf{q\cdot
R})\\
M_{x}(\mathbf{r})\sin(\mathbf{q\cdot R})+M_{y}(\mathbf{r})\cos(\mathbf{q\cdot
R})\\
M_{z}%
\end{pmatrix}
\end{equation}

The corresponding spinor wavefunction can be expressed as%

\begin{equation}
{\phi_{n\mathbf{k}}^{SS}(\mathbf{r})=%
\begin{pmatrix}
u_{nk}^{\uparrow}(\mathbf{r})e^{i(\mathbf{k-q}/2)\cdot\mathbf{r}}\\
u_{nk}^{\downarrow}(\mathbf{r})e^{i(\mathbf{k+q/}2)\cdot\mathbf{r}}%
\end{pmatrix}
}%
\end{equation}
where $u_{nk}$ are periodic in the unit cell. This theorem allows solving for
${\phi_{n\mathbf{k}}^{SS}(\mathbf{r})}$ by solving two separate Bloch
equations for $\mathbf{k\pm q/}2$ using any standard electronic structure methodology. As
mentioned in the Introduction section, Sandratskii's method is implemented in many
standard DFT packages~\cite{VASP,vasp1,elk,wien2k}.

Two \textit{caveats} are in place. First, this method is not applicable when
spin-orbit coupling is important for the energetics of the material concerned (which is not the case in
NbSe$_{2}),$ since it couples the spin-up and the spin-down components.
However, spin-orbit interaction can be added perturbatively, as it is done,
for instance, in FLEUR~\cite{FLEUR}. Second, for itinerant metals the magnetic
ground state (with an enhanced $I)$ is not necessarily an ideal spiral; it may
have amplitude variations periodic in \textbf{q}. While this does not affect
our methodology, which only exploits the properties near the instability,
$i.e.,$ near $M=0,$ it might be of interest in other cases. In particular,
even while in real life, NbSe$_{2}$ is not magnetic, the ground state in DFT-GGA is a spin density wave (SDW)~\cite{SDW}, the fact that is at least of some
academic interest, and it was claimed that the DFT ground state is
not a spiral but an amplitude-modulated SDW. If that were the case, it would
have been rather unusual for weak itinerant magnetic metals (cf. Sr$_{2}%
$RuO$_{4},$ where an amplitude SDW is nearly degenerate with the spin-spiral state, but
still loses to the latter~\cite{Kim2}). In the results section, we discuss what
happens as a matter of fact, within the framework of DFT-GGA in the case of NbSe$_{2}$ monolayer.

\begin{figure}[ht]
\centering
\includegraphics[width=0.75\linewidth]{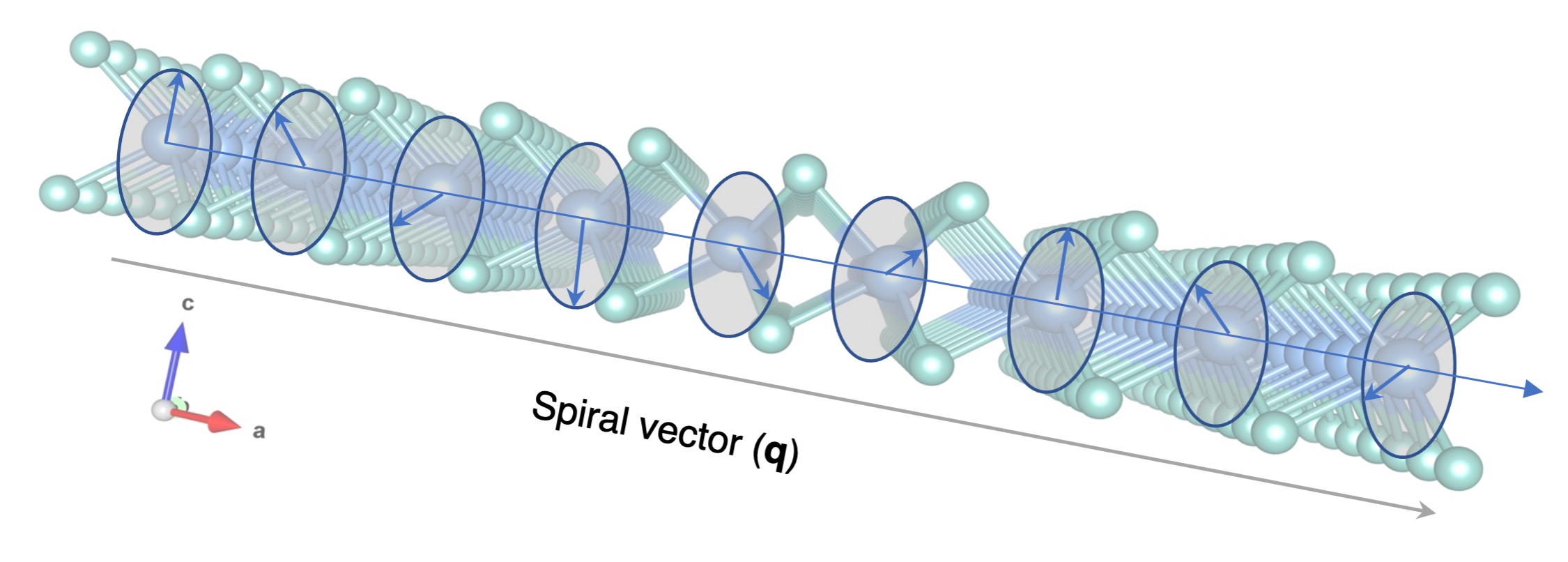}
%.
\caption{Graphical representation of the helical spin spiral
with the propagation vector
$\mathbf{q} = (\pi$/3, 0) in the monolayer
NbSe$_{2}$. Note that in non-relativistic
calculations, the energy does not depend on the orientation of the spin rotation plane.}%
\label{fig:2}%
\end{figure}

\begin{table}[ht]
\centering
\begin{tabular}{|l|l|l|l|}
\hline
Spin spiral & {E$_{ssp}$(meV)/ f. u.} & Supercell &
{E$_{ref}$(meV)/ f. u.}\\
\hline
$q_{1}$ = 0.2 & -19.818486 & 5$\times 1$  & -19.786354\\
\hline
$q_{1}$ = 0.25 & -19.817846 & 4$\times 1$  & -19.785754\\
\hline
$q_{1}$ = 0.333 & -19.816859 & 3$\times 1$ & -19.784854\\
\hline
$q_{1}$ = 0.5\hspace{1mm}(AFM) & -19.817072 & 2$\times$ 1  & -19.785280\\
\hline
\end{tabular}
\caption{\label{tab:example}Comparison of the energetics from supercell calculations in monolayer NbSe$_{2}$ with that of the spin spiral method as implemented in VASP. Note that the energetics of the $5\times 1 $, $4\times 1 $, $3 \times 1 $ and $2 \times 1 $ supercells consistently agree with that of the converged spin spiral calculations, upto a constant shift in energy of 32 meV. E$_{ssp}$, E$_{sup}$ and E$_{ref}$ = E$_{sup}$+ 32.132 meV refers to the spin spiral, supercell and reference energy respectively.}
\end{table}

\section{RESULTS AND DISCUSSIONS}

Correlated electronic phases in recently popular two-dimensional materials
such as CrI$_{3}$~\cite{CrI3_mag1,CrI3_mag2} and VI$_{3}$~\cite{VI3}, exhibit
long-range magnetic order in spite of its suppression by thermal fluctuations
 by virtue of the Mermin-Wagner theorem~\cite{Mermin,Hohenberg}. Among the
prospective quantum materials ~\cite{printed_NbSe2,proximity,MoS2,twisted}, bulk
2H-NbSe$_{2}$ has gained significant popularity due to the simultaneous observation of
superconductivity~\cite{bulkNbSe2} and charge density wave (CDW)~\cite{bulkCDW,bulkCDW2,Hirschfeld}. The CDW
transition in 2H-NbSe$\mathrm{_{2}}$ has been addressed several times computationally~\cite{bulkCDW,bulkCDW2,Hirschfeld} using the fact that the commensurate charge density wave
vector $\mathbf{q}=(1/3,0)\hspace{1mm}\mathbf{a}^{\ast}$
corresponds to a structural reconstruction within
a $3\times3$ supercell ($\mathbf{a}^{\ast}=2\pi/\sqrt{3}a$ is the reciprocal lattice
vector). NbSe$_{2}$, a layered van der Waals material, has recently inspired
the study of superconductivity in its monolayer form~\cite{ising, ising2,Mak,monoCDW,Khodas}. The proximity effect and
magnetic switching at interfaces of this material with other magnetic
monolayer TMDs~\cite{TaS2-NbSe2,Mak_arxiv} such as TaS$_{2}$, TaSe$_{2}$ and CrBr$_{3}$ warrant detailed
study of the low-energy properties in this material. The lack of inversion
symmetry in monolayers of 2H-NbSe$_{2}$ leads to a broken Kramer's spin
degeneracy and large spin-orbit (SO) splitting of the states at the momentum
$K$, and its inversion partner, $K^{\prime}=-K$, in the Brillouin zone. The
magnitude of SO-splitting in the monolayer is much larger than the
superconducting order parameter~\cite{Darshana, Mak_arxiv}. The combination of SO-coupling and broken
inversion symmetry results in locking of the pseudospins at the points $K$ and
$K^{\prime}$ to be parallel to the $c$-axis of the monolayer. As a result of
time-reversal symmetry, the pseudospins at the $K$ and $K^{\prime}$ points are
antiparallel, with degenerate energies. The ensuing novel phenomenon was aptly
named \textquotedblleft Ising superconductivity\textquotedblright~\cite{Mak,ising,ising2}. In quantum confined monolayers, screening is significantly reduced compared to bulk,
leading to enhancement of electronic correlation. In DFT, this leads to a
magnetic instability in the undistorted monolayer, which is remedied either
by the formation of a charge density wave, or through quantum fluctuations.
\newline
\par

In this section, we elucidate the results pertaining to interesting magnetic
phases calculated for the monolayer 1H-NbSe$\mathrm{_{2}}$. As shown in Fig.
1, the spin spiral calculations~\cite{Sandratskii,Sandratskii2} were performed for this systems for various
spiral vectors $\mathbf{q}$ over a fine momentum grid across the entire irreducible Brillouin zone.
Note that the spiral vectors are defined so that the magnetic moment
associated with the atomic positions in the atomic lattice have no amplitude
along the longitudinal direction of propagation of spiral~\cite{Sandratskii,Sandratskii2}, hence excluding
magnetic patterns with nonzero net magnetization. Thus this arrangement
corresponds to either helical or cycloidal spin spiral (which have, in the
absence of spin-orbit, the same energy). For test purposes, we have
performed supercell calculations for selected spiral wave vectors.
Specifically, we have generated supercell that allowed us to calculate
commensurate spirals with $\mathbf{q}=(q_{1},0),$ where $q_{1}=\frac{1}{5}$,
$\frac{1}{4}$, $\frac{1}{3}$ and $\frac{1}{2}$. The comparison of total
energies per formula unit for the different spin orientations as obtained from
the calculations are presented in Table I. Apart from a constant energy shift
of 32.13 meV the spin spiral calculations fully agree with those in the
supercells. Either way, we recognize the DFT ground state to be a spiral with $\mathbf
{q}\approx {(0.2,0)}$. A previous investigation of magnetic ordering in the
monolayer NbSe$_{2}$ suggested~\cite{feng} the lowest energy phase to be
nearly collinear antiferromagnetic (without a CDW) corresponding to the 4
$\times$ 1 supercell. However, our calculations find this state to be still
higher in energy than the $({\frac{1}{5},0)}$ spiral.

\begin{figure}[ht]
\centering
\includegraphics[width=0.75\linewidth]{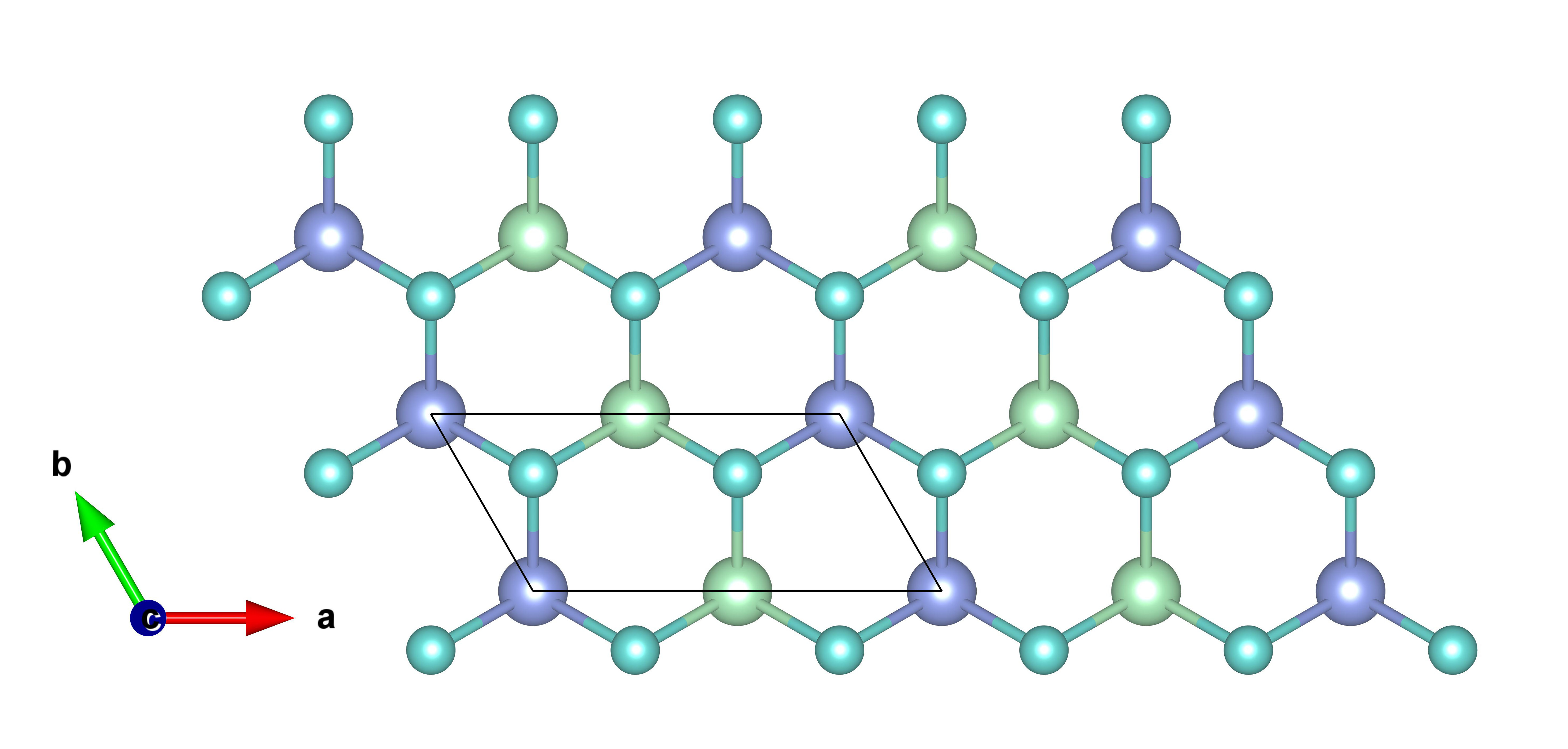}
%Here is
\caption{The crystal lattice structure of monolayer NbSe$_{2}$ as observed from the top ($c$-direction). While various magnetic ordering of the material are explored, here we
show the prototypical up and down sublattices in the antiferromagnetic configuration.}%
\label{fig:3}%
\end{figure}In Figure 2, we display the lattice structure of a single layer of
NbSe$_{2}$ as viewed from above (along the $c$-direction). Note that the Nb atoms are bonded to the adjacent Se
atoms in a trigonal prismatic coordination. In order to study the magnetic phases in 
monolayer NbSe$_{2}$, constrained fixed spin moment (FSM) calculations were
performed where the magnetic moment of the monolayer is varied and the energy
difference of the magnetic and nonmagnetic states is calculated. We then fit the calculated total energy as a function
of magnetization:
\begin{equation}
E(M)=a_{0}+a_{1}M^{2}+a_{2}M^{4}+a_{3}M^{6}+....
\end{equation}
and use Eq. \ref{dft} to determining the uniform spin susceptibility
${\chi_{DFT}(\mathbf{q}=0)}$ from the fitting parameter $a_{1}$.
From the FSM calculations at  $U_{eff}=0$ we find ${\chi_{DFT}(\mathbf
{q}=0)=6.87\times10^{-4}}$ {emu/mol.}

Next, we perform FSM calculations for different values of $U_{eff}$ (Fig.
\ref{fig:5}). At some values of $U_{eff}$ (in this plot, $U_{eff}=0.7$ eV) the
curve $E(M)$ has two minima, $M=0$ and another one at a finite moment. One of these minima corresponds to the ground state, and the other to a metastable solution~\cite{Khomskii}. Either way, for the purpose of determining the susceptibility, we
need to know the behavior at small $M.$ 

From the full $E(M)$ curve at each $U_{eff}$ we can find $a_{1}$,
and we observe that, at the critical value $U_{c}$ = 0.918, $a_{1}$ becomes zero
and the uniform $\mathbf{q}$ = 0 state becomes unstable against ferromagnetism
(Fig. \ref{fig:5} (b)). Comparing the already known value of ${\chi_{DFT}%
(\mathbf{q}=0)}$ with the $U_{c}$ = 0.918 and using Eq. \ref{Uc}, we can find the
constant $\kappa$ in that equation, $\kappa=1.586\times10^{3}$ mol/emu.

\begin{figure}[ht]
\centering
\includegraphics[width=.5\linewidth]{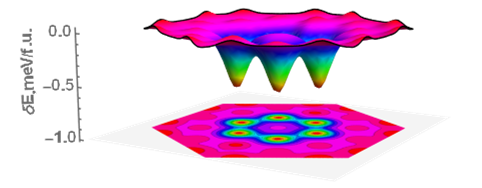}
\includegraphics[width=.47\linewidth]{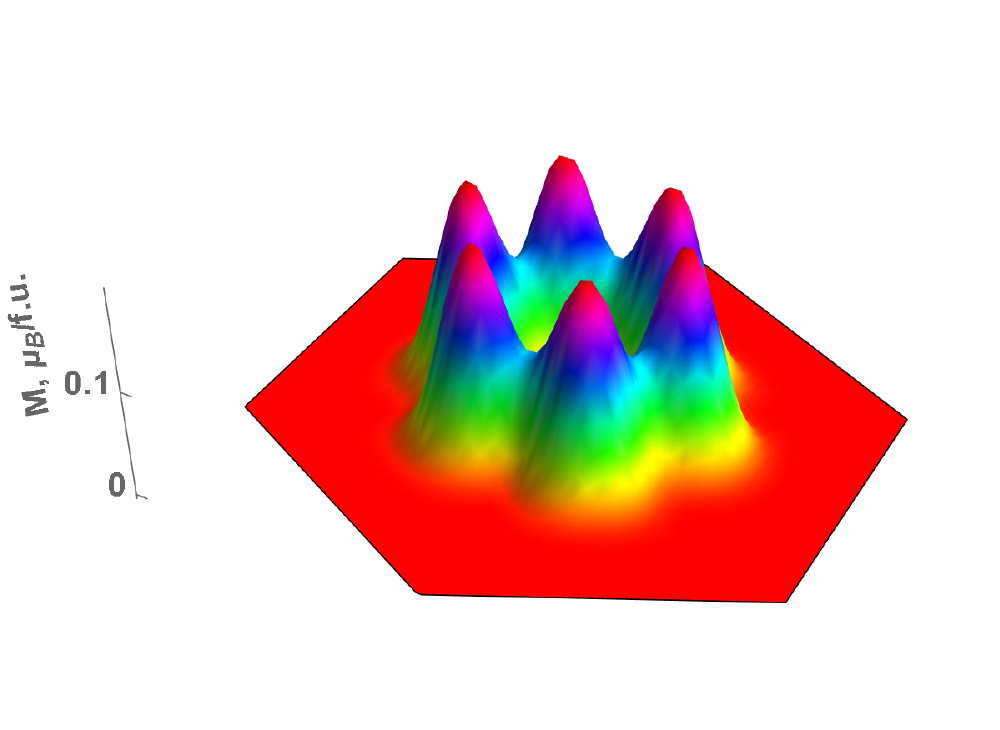}
\caption{
(a) Energies of the spin spiral states across the full Brillouin zone (BZ) of monolayer NbSe$_{2}$; outside of the narrow regions near  $\mathbf{q}=(0.2,0)$ the spiral calculations collapse, so the energy difference is zero (apart from some numerical noise introduced by the plotting software). (b) Same, for the magnitude of the magnetic moment calculated for the spin spiral.}%
\label{fig:4}%
\end{figure}

\begin{figure}[ht]
\centering
\includegraphics[width=0.75\linewidth]{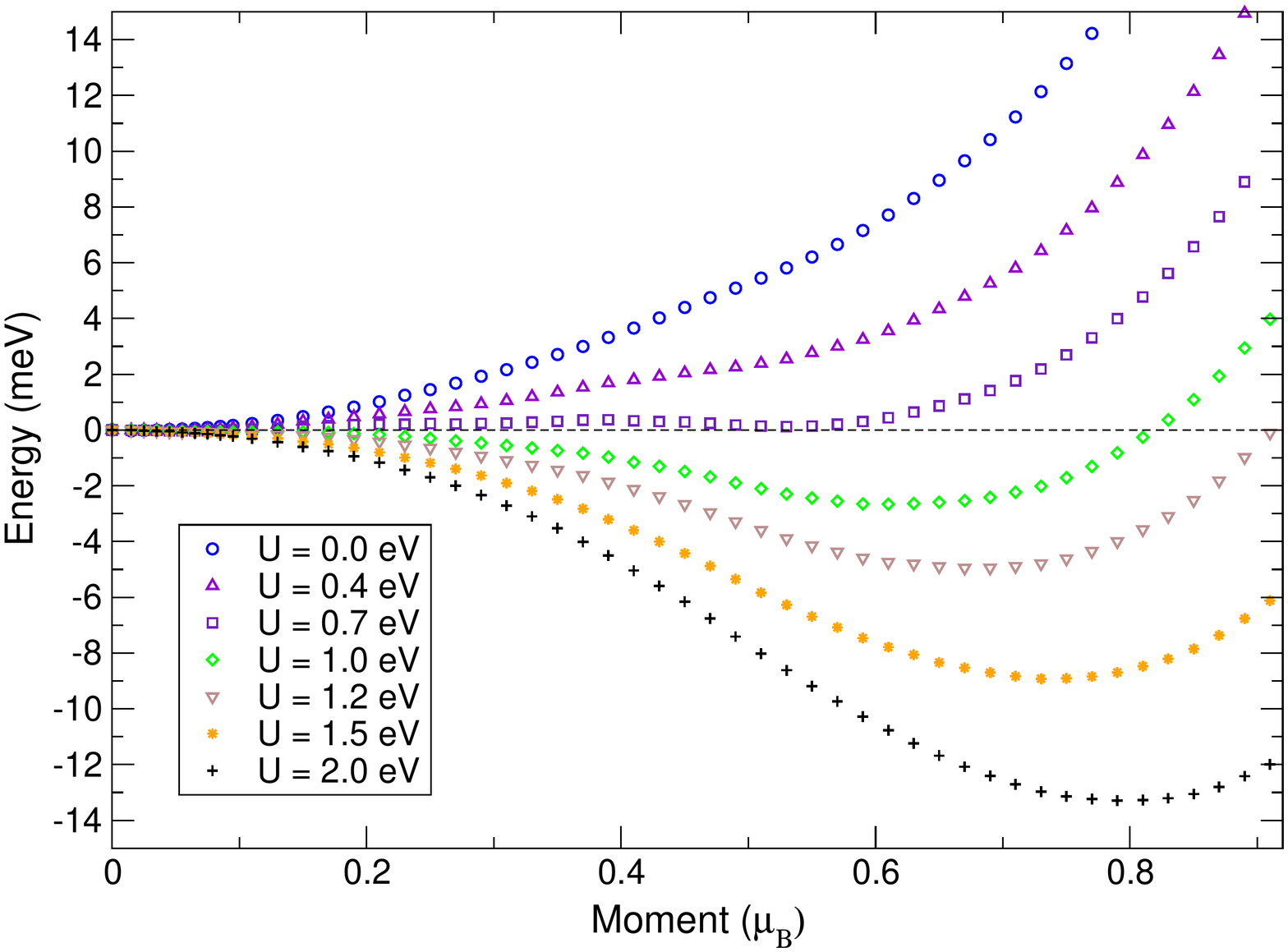}
\includegraphics[width=0.75\linewidth]{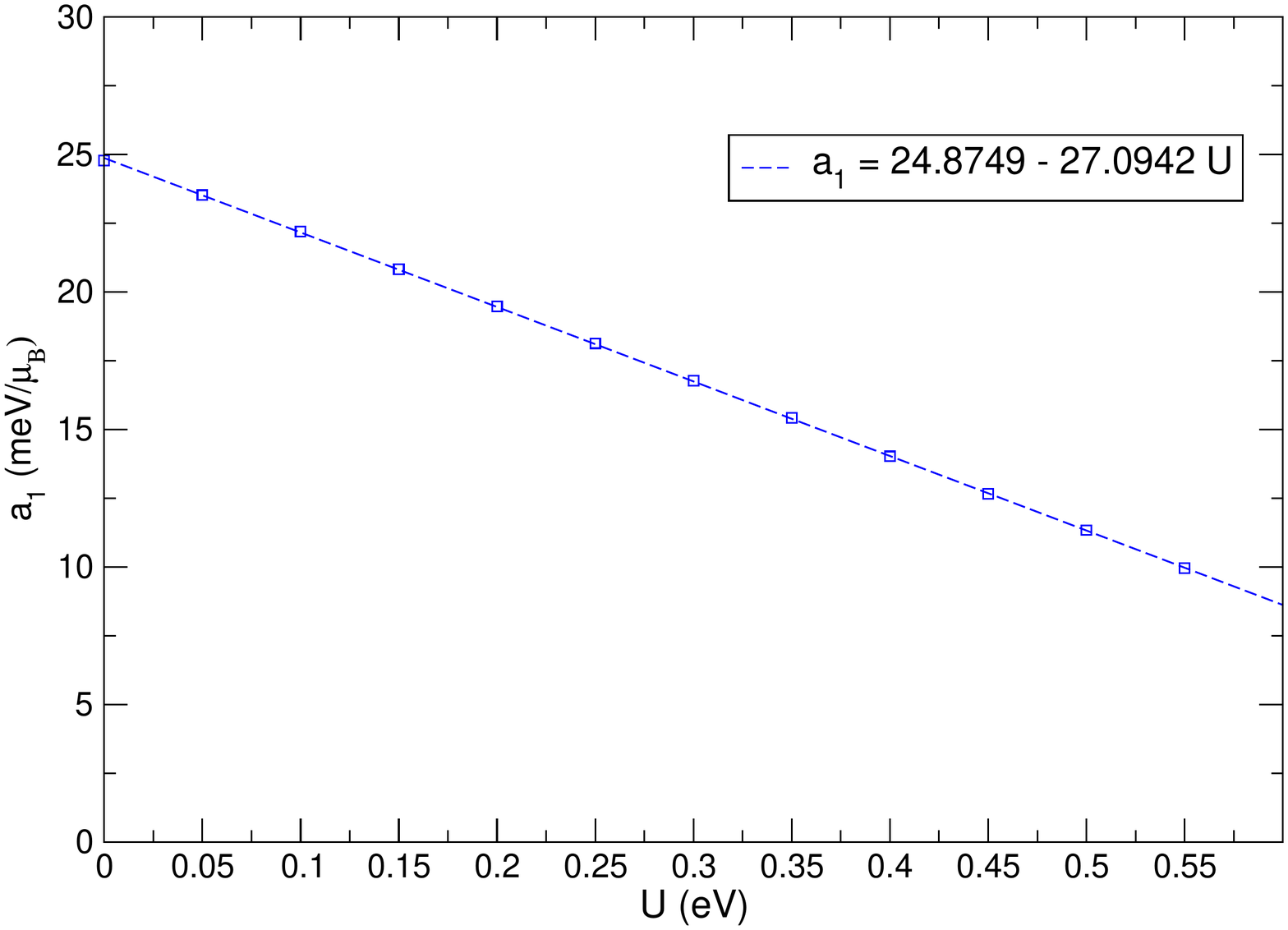}
\caption{(a) Fixed spin moment 
calculation (FSM) for the uniform magnetization $\mathbf{q}$ = 0, for various effective
Hubbard interaction values ranging from $U_{eff}=$ 0 to $U_{eff}$= 2.0 and (b) {
determination of the critical value $U_{eff}$ = 0.918 from the slope of $a_{1}(U_{eff})$ 
from the magnetic instability condition}.}%
\label{fig:5}%
\end{figure}

Now we are ready to address the spiral states. The calculated energy and magnetic
moment at a uniform k-point mesh of spiral vectors $\mathbf{q}$ are presented in Fig. \ref{fig:4}.  Fig. \ref{fig:4}
(a) elucidates the energy spectrum obtained from accurate spin spiral calculations presented as a color map for the
entire 2-D hexagonal Brillouin zone. Note that the spin spiral calculations with
spiral vectors $\mathbf{q}={\frac{1}{5}\hspace{1mm}\mathbf{a}^{\ast}}$ correspond to the
5$\times$1 supercell of monolayer NbSe$_{2}$. Our calculation indicates a sharp
energy minimum at this spiral vector, $\mathbf{q}={\frac{1}{5}%
\hspace{1mm}\mathbf{a}^{\ast},}$ where the spiral magnetic moment also exhibits a
maximum. That is to say, even though the actual material is not magnetically ordered, it is liable to have strong spin fluctuations
at and near $\mathbf{q}_c=(0.2,0)$. The calculated DFT magnetic moment [Fig. \ref{fig:4} (b)] is nonzero in a narrow region near  $\mathbf{q}_c$.  Our supercell calculations confirm the
existence of magnetic instability at this particular wave vector.
\newline

So far we have discussed unenhanced and unrenormalized DFT calculations. Next, we report energies from spin spiral calculations with an artificially
enhanced Hubbard interaction.~\cite{Dudarev}

Available electronic structure codes~\cite{VASP,elk,FLEUR} do not allow FSM calculations for nonzero spiral vectors. Instead, in order to find the critical values
$U_{c}(\mathbf{q})\ $corresponding to the onset of an instability, we start
calculations from a very small initial magnetic moment of 0.01 $\mu_{B}$ and
monitor whether the magnetization will remain on the level of computational
noise, or converge to a finite magnetic moment. Starting from a sizeable
$M_{0}$ for some spiral vector actually leads to a magnetic instability with a
finite self-consistent $M,$ even though the $m=0$ state remains metastable and
the susceptibility finite. Of course, such solutions are of no use for determining
the susceptibility.

\begin{figure}[ht]
\centering
\includegraphics[width=.5\linewidth]{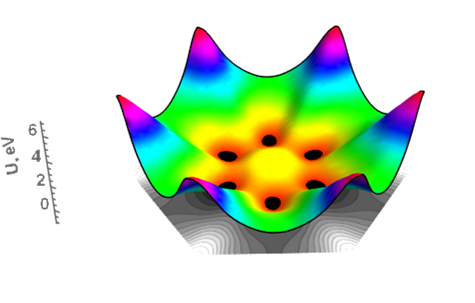}
\includegraphics[width=.3\linewidth]{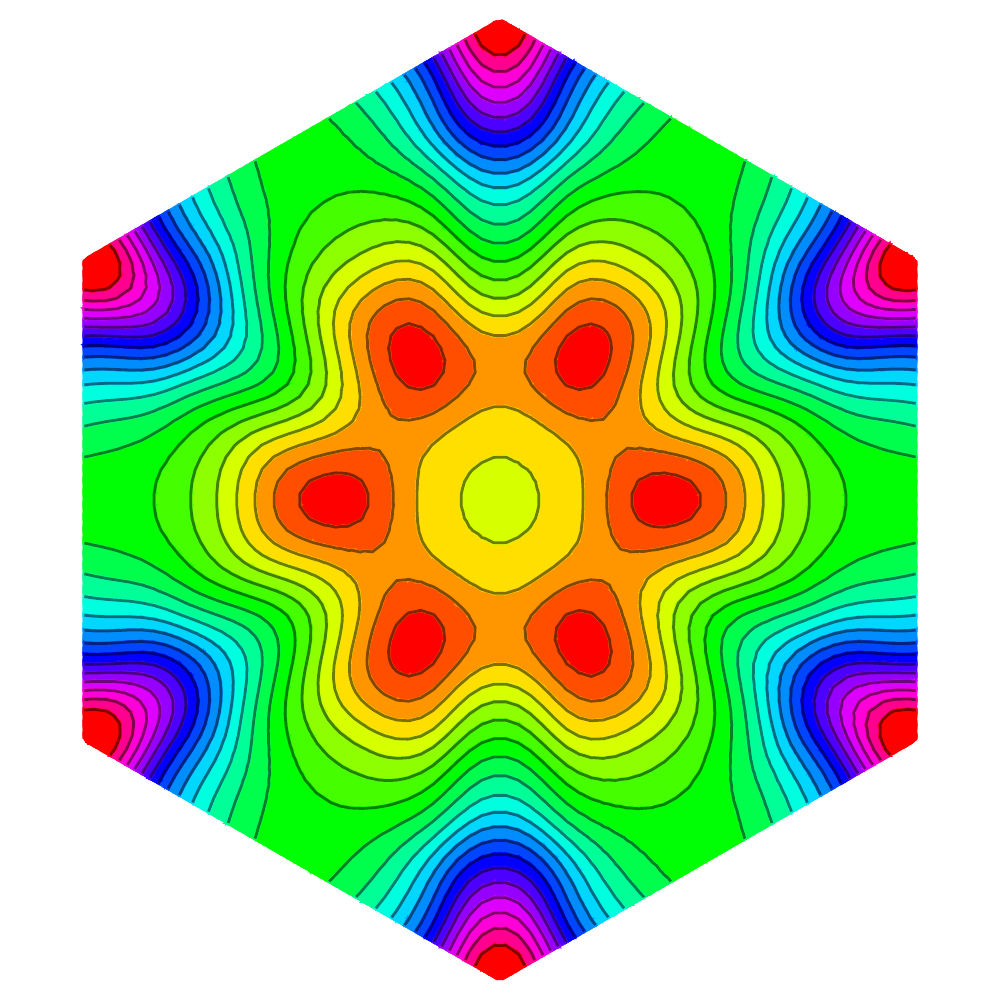}
\caption{(a) Critical value of effective Hubbard interaction $U_{eff}$ defining the magnetic instability as a function of the spiral vector $\mathbf{q}$ across the entire Brillouin zone of 
monolayer NbSe$_{2}$ and (b) the corresponding contour plot as viewed from above.}%
\label{fig:6}%
\end{figure}

For spiral vectors close to $\mathbf{q}=(0.2,0)$ 
the nonmagnetic
solution is unstable even for $U_{eff}=0.$ In those cases,
we were adding a
\textit{negative} $U_{eff}.$ While negative values of $U_{eff}$ are
nonphysical, they provide us with an instrument to extract the unrenormalized
DFT susceptibilty $\chi_{DFT}(\mathbf{q})=1/\kappa U_{c}(\mathbf{q}),$ which, in
those cases, is negative. Fig. \ref{fig:6} (a) shows $U_{c}$ as a function of
$\mathbf{q}$ in the 2D hexagonal Brillouin zone. It varies from $-1.0$ eV at
$\mathbf{q}=(0.2,0)$ to 6.0 eV at the Brillouin zone edge
$K$. We do not plot $\chi_{DFT}(\mathbf{q})$, since it is just inversely
proportional to $U_{c}(\mathbf{q})$ plotted in  Fig. \ref{fig:6} (a).

Our next step is to renormalize the DFT spin susceptibilities in the spirit of Moriya's theory~\cite{Moriya, Ortenzi}. As we already know, $\chi_{DFT}(\mathbf{q}=0)=6.87\times10^{-4}$
emu/mol., while the non-interacting susceptibility $\chi_{0}=0.872\times10^{-4}$ emu/ mol., from the density of states at the Fermi level $N(0)=2.7$ states/f.u.
and the DFT Stoner factor can be calculated to be  $I=0.646$ eV/f.u. Now, applying Eq. 12 to the magnetic instability corresponding to
spiral vector $\mathbf{q}=0$ and critical effective Hubbard interaction $U_{eff}=0$
yields $\kappa=1.56\times10^{3}$ (mol/emu)/eV ($\mathrm{Tr}{(D\cdot D)/{N_{F}^{2}}}=0.1$ in Eq. \ref{eq:eq3}).

\begin{figure}[ht]
\centering
\includegraphics[width=.5\linewidth]{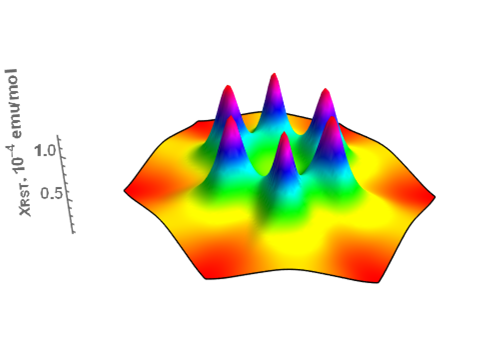}
\includegraphics[width=.3\linewidth]{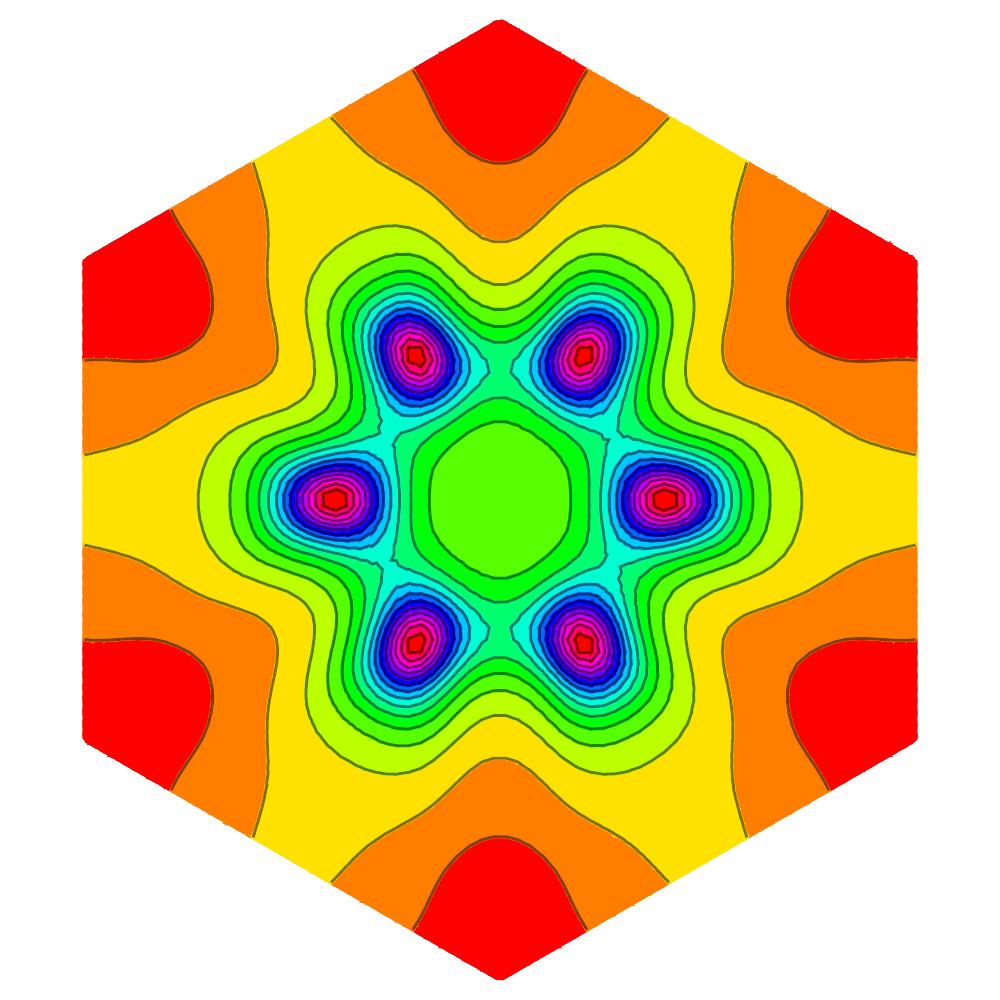}
\caption{The figure shows (a) the fluctuation-renormalized spin susceptibility as a function of the spiral vector $\mathbf{q}$ across the entire Brillouin zone of monolayer NbSe$_{2}$, and (b) the corresponding contour plot, of $U_{eff}$, as viewed from above.}%
\label{fig:6a}%
\end{figure}

Following the formalism for Reduced Stoner Theory~\cite{Petukhov}, we introduce the
fluctuation-induced Moriya factor $\alpha,$ so that $I_{eff}=\alpha
I,\alpha<1$. Using
\begin{equation}
\chi_{RST}^{-1}(\mathbf{q})=\chi_{DFT}^{-1}(\mathbf{q})+(1-\alpha)I\label{eq:RST}%
\end{equation}
we can determine $\alpha$ by comparing Eq. \ref{eq:RST} with the experimental
spin susceptibility, if the latter is available. As of now, the experimental
spin susceptibility has been measured only for the bulk sample of NbSe$_{2}$~\cite{expt_chi}.
The bulk experimental and first principles spin susceptibilities for
$\mathbf{q}$ = 0 are $\chi_{expt}=3\times10^{-4}$ emu/mol. and $\chi
_{DFT}=4.28\times10^{-4}$ emu/mol.
Assuming the contribution from spin fluctuations in monolayer to be the same, we use $\alpha$ = 0.891. Utilizing this $\alpha$ and
$\chi_{DFT}(\mathbf{q})$, we obtain the fully renormalized $q$-dependent spin
susceptibility for monolayer NbSe$_{2}$, shown in Fig. \ref{fig:6a} as a function of the spiral vector
$\mathbf{q}$. Note that we observe two maxima, a weak peak around
the $\Gamma$ point  and
the principal set of peaks around six points equivalent to $\mathbf{q}=(0.2,0)$. 

\section{Summary and Conclusions}

In summary, we have designed a protocol to estimate both DFT~\cite{noncolin, GGA,Kresse} and
fluctuation-renormalized~\cite{Petukhov, Ortenzi, Ron} (in the spirit of Moriya's theory) spin
susceptibility, especially well suited for materials close to a magnetic
instability, but not surpassing it. The protocol does not require
linear-response calculations~\cite{Savrasov}, nor explicit accounting for fluctuations~\cite{Antropov, Kutepov}, as it
is done, for instance, in dynamical mean field theory. It is based on the
capability to tune a material's propensity to magnetism by including a variable LDA+U correction~\cite{Dudarev} (even in a weakly correlated material), and then
reverse-engineering the standard RPA formula~\cite{Pines}.

The formalism includes two \textit{a priori }unknown constants, assumed to be
$\mathbf{q}$-independent, one of which can be fixed by a comparison with the
fixed spin moment calculations at $\mathbf{q}=0,$ and the other by a comparison
with the experimentally observed uniform spin susceptibility. The capabilities
to perform  FSM calculations at $\mathbf{q}=0,$ and self-consistent spiral
calculations at an arbitrary $\mathbf{q}$ are built-in within most standard DFT codes.

We apply this procedure to a 2D Ising superconductor, monolayer NbSe$_{2}$. We
find very strong antiferromagnetic spin fluctuation at and near $\mathbf
{q}=(0.2,0),$ indicating that the structure of spin fluctuations in the
momentum space in this superconductor is more complicated that previously
thought of. These findings have direct ramifications for the structure of the
superconducting order parameter in monolayer NbSe$_{2},$ especially on the
degree of the singlet-triplet mixing~\cite{Mak_arxiv}. These ramifications will be discussed in a separate publication.

\section{Methods}

\subsection{Computational Methods}

We have employed the generalized gradient approximation (GGA) for the exchange
correlation functional and the projector augmented wave method 
as implemented within the Vienna Ab initio Simulation Package (VASP)
Code~\cite{VASP, vasp1}. The VASP electronic structure code does not adopt any particular approximation to either
the charge or magnetization density, or electronic potential, thereby
allowing for interatomic as well as intra-atomic noncollinearity of the spin
density. Calculations using Generalized Gradient Approximation (GGA) and
Hubbard $\mathrm{U}$ type corrections for localized d electrons (GGA+U) were performed
utilizing the Dudarev approach~\cite{Dudarev} where the difference ${(U-J)}$ is
incorporated as an effective term ${U_{eff}}$. The Nb pseudopotential
in our calculations includes the 4d and 5s electrons in the valence bands (inclusion of the latter proved to be quite important). The
single-particle wave functions were evaluated using a plane-wave
energy cutoff of 600 Ry. The spin susceptibility $\chi
(\mathbf{q)}$ was evaluated on the 6$\times6$ mesh in the irreducible wedge of
the Brillouin zone.

\section{DATA AVAILABILITY}

The datasets generated during and/or analyzed during the current study are available from the corresponding author on reasonable request.

\bibliographystyle{unsrtnat}
\bibliography{full-spin_sus.bib}
%Produces the bibliography via BibTeX.
%\section{Acknowledgements}

\begin{acknowledgments}
The authors acknowledge financial support for their research from the Office of Naval Research through Grant N00014-20-1-2345. The authors also acknowledge the Department of Defense (DoD) Major Shared Computing Resource Center at Air Force Research Laboratory (AFRL) and the National Energy Research Scientific Computing Center (NERSC) for high-performance computing facilities for significant portion of calculations performed in this work. We thank V. P. Antropov for useful discussions.
\end{acknowledgments}

\end{document}